# Comments on "Low Q nuclear fusion in a volume heated mixed fuel reactor" [6] by H. Ruhl and G. Korn (Marvel Fusion, Munich)

K. Lackner, R. Burhenn, S. Fietz, A. v. Müller

Max Planck Institute for Plasma Physics, 85748 Garching, Germany

Abstract

We comment on the note "Low Q nuclear fusion in a volume heated mixed fuel reactor" by Ruhl and Korn in which hydrodynamic life-time considerations are included in estimates of ignition energy for uncompressed fusion fuel targets. For the case of DT fusion, the authors arrive at a required hot-spot energy of 1MJ. We point out that their $Q = 1$ estimate is not relevant for transition into a thermonuclear fusion regime, and that it is based on DT implanted in a boron-proton matrix without accounting for all consequences. Applying the proper corrections leads to an increase in the needed laser energy to initiate useful fusion power production, even for DT, into the GJ range.

The aim of Marvel Fusion is the use of p-$^{11}$B fusion reactions, which would require already according to Ruhl and Korn's optimistic estimate, a hot-spot energy of 1 GJ. If this were to be provided for by a mixed pBDT fuel or a staged explosion scheme (rather than by laser deposition), it would imply a high associated production rate of fast neutrons, and the need for tritium breeding, and push the energy produced by a single laser pulse into non-manageable dimensions.

## 1. Introduction

Marvel Fusion aims at the near-term commercial utilization of proton-boron (p-$^{11}$B) fusion using advanced laser system. Distinguishing features of their approach are the utilization of nano-structured targets, of ultra-fast (100 fs), ultra-powerful lasers as drivers, and the renunciation of pre-compression of the fuel target. In previous notes, Ruhl and Korn [1,2] described a code system for analysing laser-plasma interaction, predicting the efficient creation of fast ions, and gave some general features of the invoked nuclear reactions.

Nuclear fusion schemes that do not provide for confinement by externally applied magnetic fields or gravity are limited in their lifetime by the disintegration of the fuel assembly, which can proceed on a time-scale too short to allow useful energy production [3]. In previous preprints, the authors ignored this most basic limitation. As this competition depends only on the presence of kinetic energy and the well-known cross-sections for the relevant nuclear reactions, it holds - in only slightly differing form - for all kinds of thermal and fast ion-driven reactions. We pointed out in [4,5] that these constraints rendered the estimates given by Ruhl and Korn irrelevant and clarified also that self-generated magnetic fields could not contribute to confinement due to the Virial Principle.

Ruhl and Korn have recently provided several updates to their proposal, see References [6] and [7]. They differ in title, assumptions and models used, so that we will comment on them in two separate notes. The first, original version of their manuscript, Reference [6], contains a rudimentary hydrodynamic model for the fuel-disintegration, which is the focus of our first comments and will be discussed below.

## 2. Estimates of hot spot requirements for fusion energy production

In the absence of confining external magnetic fields, fusion energy production is limited by the finite life-time of the fuel assembly due to its finite size. This is the result of the competition between energy production and disintegration of the target and gives rise to the well-known criteria on the required line-density $(\varrho R)_{fus}$ of the initially heated fuel for ignition, and of the fuel assembly for the burn-up fraction. The scaling of the number of particles in a volume $(\sim \varrho R^3)$ with density for a fixed line density $\left(\varrho R^3 \sim (\varrho R)_{fus}{}^3/\varrho^2\right)$ leads to a quadratic decrease of the energy required in this hot spot with increasing density, and motivates the about 1000fold compression assumed in conventional inertial fusion concepts [3].

Reference [6] includes such "hydrodynamic" finite life-time effects of the fuel assembly, relating them to the transit time of an ion-acoustic wave - a standard procedure, proven to lead, when applicable, to correct scalings and ball-park parameters. The fuel is assumed to consist of a boron-proton matrix implanted with a variable amount of DT, with D-T and p-[11]B fusion reactions contributing to the energy production. Performance parameters given by the authors that can be compared with standard literature [3,8] are the fusion yield $Q$ and the volume ignition (hot spot) energy $E_i$. This kind of analysis results in a value of $\varrho R$ and of $E_i$ as function of temperature required to achieve a desired $Q$-value. The hot-spot energy (at least in the low yield regime) would have to be provided for by external heating, i.e., ultimately by net deposited laser power in this region to produce the fusion reactions.

Without pre-compression, Fig. 4 of their paper predicts a required hot-spot energy $E_i$ for the most favorable case of pure DT of about 1 MJ - higher than any energy quoted in the previous publications [1] and [2] - but surprisingly still in the ballpark of the total laser energies involved in the recent, successful Livermore experiments [8]. Of course, in the latter case this energy had to be provided primarily for *compressing* the fuel assembly by a large, but conventional and rather "slow" laser system, but in the Marvel proposal by a much more advanced sub-100fs laser system.

*However, this estimate by Ruhl and Korn has to be corrected upward by more than 3 orders of magnitude to correspond to an equivalent thermonuclear burn situation and to give a truly relevant estimate for the size of the laser system needed to achieve physically or practically relevant results.*

The 1 MJ estimate by Ruhl and Korn is derived, by aiming at $Q$=1, for a DT fuel according to figure 4 in Reference [6]. However, for DT this is far from yielding physically or practically significant effects of self-heating. An energy gain factor of $Q$=1 is universally labelled "breakeven", and not associated with "ignition", as is done by the authors. At $Q$=1 it is *not* true that, as stated in Reference [6], "due to the combined effects at ignition the reactor rapidly transitions into thermonuclear fusion". This is due to the fact that in DT (in contrast to p-[11]B) 80% of the fusion energy appears in the form of neutrons and only 20% is available for plasma self-heating through α-particles. The relevant value for describing entry to the thermonuclear burn regime is therefore $Q$=5. This has dramatic consequences, as - prior to the onset of self-heating - the accepted scaling-relations of the required hot-spot energy give a dependence on $Q$ according to $E_i \sim Q^3$ (see equation 1 in Reference [6]), implying an increase by a factor of 125 (!) for Q=5. Moreover, Marvel's basic scheme for laser-plasma coupling is based on a nano-structure of the boron containing target, which should eliminate also the need for cryogenic fuel. This nano-structure implies, however, also the presence of boron atoms which contribute - in the low-energy scenarios - nothing to energy production, but rather to the specific heat of the fuel assembly. Thus, the most favorable but realistic scenario in Fig. 4 of Reference [6] is their case labelled b3, which requires, a further ten-times larger hot-spot energy (10 MJ for $Q=1$).

The authors also invoke a further benefit from the boron-matrix, even in the pure DT fusion case, by using the ion acoustic speed $u^s$ of boron in the estimate of the disintegration time, benefiting from the increased inertia. Their expression, however, holds only for singly-ionized boron, which is not a good approximation at the 30 keV temperatures of the reference case[i]. For fully ionized boron, the ion-acoustic speed would be a factor of $\sqrt{3}$ larger, which would increase the estimate of $E_i$ by a further factor of 5.2. Multiplied together, these factors suggest that in reality, without compression a deposited laser hot-spot energy of 6500 MJ, rather than 1 MJ would be needed! This corrected result is in a ballpark with the expression 4.24 given in Reference [3], which - for true (isochoric) ignition of pure DT at solid state hydrogen density - implies the need of a hot spot energy of 14 GJ.

The grossly overoptimistic nature of the estimate of Ruhl and Korn for the laser energy required to see effects to be associated with thermonuclear burn is actually also documented in their own conclusions, where they apply their estimates for $E_i$ also to a case *with* fuel pre-compression. They derive a requirement for 10 J - more than a factor of thousand smaller than actually needed in the successful Livermore experiments [8].

At first glance, even the most optimistic results of Ruhl and Korn would appear to already rule out proton-boron fusion. Among their cases studied in Fig.4 of Reference [6], only case b5 corresponds to dominating proton-boron fusion. It requires a starting hot-spot energy on the order of 1 GJ, clearly out of the range of direct laser initiation. But of course, in a mixed or staged burn scheme, this energy could be provided by a preceding stage of high-Q DT burn, which would, however, be associated with a corresponding production of fast neutrons, and largely annul the stated advantages of the p-[11]B reaction chain (no fast neutrons, meaning no material damage in the reactor walls and no need for tritium breeding). To lay claim to the advantages of p-[11]B-fusion, this second burn stage would have to multiply this 1GJ starting energy again by a large factor, raising the single-explosion energy most probably far beyond commercially usable dimensions.

## 3. Applicability of the analysis of Ruhl and Korn to their described configurations

In fact, however, the above standard-inertial fusion model may not even be applicable to the fuel and ignition configurations envisaged by the proponents, delivering systematically much too optimistic predictions for their configuration.

The simple model used above by Ruhl and Korn treats the burn region as a continuous, single region. This is justified only, if all the laser-energy were to be deposited and converted into plasma energy within a radius $(\varrho R)_{fus}/\varrho$. Laser spots at distance from each other larger than $(\varrho R)_{fus}/\varrho$ do not interact with each other during the fusion time, and have to satisfy separately the ignition hot-spot energy requirement. For the most favorable case of the authors (b3 of Figs.

---
[i] The authors state that the fusion reaction time will be long compared to the Spitzer electron-ion equilibration time, so that $T_e \approx T_i$.

3,4 in Reference [6]) the diameter of the ring formed by the light-grey areas in their Fig. 1 should therefore not exceed 1 mm[ii].

If self-heating effects and ignition are definitely not aimed at, fusion energy could be produced also by a once-through process of beam fusion in a relatively cold environment. The process would thus appear to be linear for each fast ion, with no synergy between single laser hot-spots required. However, efficient beam-target fusion requires high enough electron temperatures (in excess of about 5 keV for DT and >100 keV for p-$^{11}$B reactions) to avoid excessively fast slowing down of the ions. This requirement for electron heating introduces a similar ($\varrho R$) - criterion and the need to place laser-spots in extreme proximity of each other so as to profit from their synergy in raising the electron temperature, as we have outlined in [5]. Note that although the conventional slowing-down expressions for a plasma do not hold in a neutral medium, the electron friction given by the well-tested Bethe-Bloch [9,10] formula in the latter case has similar consequences: for 600 keV protons in boron, it gives a slowing down rate approximately equal to that on the electrons in a 0.5 keV plasma, and is therefore much too high to infringe upon the known requirement for >100 keV electron temperatures.

## 4. Possible ignition scenarios involving compression

As mentioned by the authors also in the conclusions, compression would drastically reduce the required energy for ignition. However, the configuration outlined in Figure 1 would not do so because unless contained within the radius $(\varrho R)_{fus}/\varrho$ necessary for ignition, it would give rather rise to a disintegration of the fuel ensemble before an adequate number of fusion reactions could take place. Similar concerns also preclude the possibility to use the energy of a DT explosion to ignite a separate p-$^{11}$B fuel assembly: nuclear explosions tend to propagate spherically outward and are therefore intrinsically unsuitable to produce compression. The conceivable way out is the scheme of the hydrogen bomb: using neutron or gamma radiation produced by a first stage of DT fusion to heat up a radiation cavity leading to compression and ignition of a p-$^{11}$B capsule following the indirect-drive concept. Given the high required ignition energies and the inherent inefficiencies of the intermediate steps, this would, however, inevitably exceed manageable dimensions.

## 5. Conclusions

In their recent note, Ruhl and Korn include for the first-time hydrodynamic considerations into the analysis of their fusion proposal, which renounces fuel pre-compression to exploit the intrinsic features of their laser-plasma coupling scheme. Such an analysis gives a lower estimate of the required laser powers and energies for producing a given fusion yield $Q$. Even when using the most favorable DT reaction, according to their model more than 100 MJ energy (rather than the 200 kJ quoted in Refs. [1] and [2]) would have to be delivered to the hot spot by a very advanced laser system to achieve a fusion yield of $Q$=5, frequently identified with ignition (as only 1/5 of the fusion energy of DT is available for plasma heating), and still far below the recent achievements of Livermore. To account consistently for the presence of boron (needed in their scheme also to contain the DT fuel and form the nano-structure) this estimate would have to be increased further by more than an order of magnitude.

For p-$^{11}$B fusion, an energy of about 1 GJ would have to be initially provided to a hot spot for p-$^{11}$B fusion to contribute a similar amount to plasma heating. This starting energy would not

---

[ii] The present note does not give any absolute dimensions of the proposed configurations. An earlier version of Ref. [2] made available to us, showed, however, a similar configuration with a stated diameter < 1 cm.

need to be provided for by a laser-system, but could, in principle, be produced by a mixed fuel or a two-stage assembly by DT-fusion. However, this large DT energy per pulse would render the claim of aneutronic fusion obsolete, and push the single pulse energy into non-manageable dimensions.

As final remark, as we have previously pointed out, nuclear fusion is too slow a process to profit from ultrafast laser systems, unless strong pre-compression of the target is involved. This statement is actually born out also by the analysis of the authors, who find the required ignition power to increase inversely proportional to the laser pulse length (equ. 5 of [6]). This implies that shortening the laser pulse does not reduce the required energy but increases greatly the demand on the laser technology.